\documentclass[journal,twoside,web]{ieeetran}
\usepackage{generic}
\usepackage{cite}
\usepackage{amsmath,amssymb,amsfonts}
\usepackage{algorithmic}
\usepackage{graphicx}
\usepackage{textcomp}
\usepackage[binary-units=true]{siunitx}
\usepackage[hidelinks]{hyperref}
\usepackage{nameref}
\def\BibTeX{{\rm B\kern-.05em{\sc i\kern-.025em b}\kern-.08em
    T\kern-.1667em\lower.7ex\hbox{E}\kern-.125emX}}
\begin{document}
\newcommand{\RNum}[1]{\uppercase\expandafter{\romannumeral #1\relax}}
\title{A lightweight hybrid CNN-LSTM model for  ECG-based arrhythmia detection}
\author{Negin Alamatsaz, Leyla s Tabatabaei, Mohammadreza Yazdchi, Hamidreza Payan, Nima Alamatsaz and Fahimeh Nasimi
\thanks{Negin Alamatsaz  is with the Department of Biomedical Engineering, Faculty of Engineering, University of Isfahan, Isfahan, Iran (e-mail: n.alamatsaz@eng.ui.ac.ir). }
\thanks{Leyla s Tabatabaei is with the Department of Biomedical Engineering, Faculty of Engineering, University of Isfahan, Isfahan, Iran (e-mail: l.tabatabaei@eng.ui.ac.ir). }
(Corresponding author: Mohammadreza Yazdchi.)
\thanks{Mohammadreza Yazdchi is with the Department of Biomedical Engineering, Faculty of Engineering, University of Isfahan, Isfahan, Iran (e-mail: yazdchi@eng.ui.ac.ir).}
\thanks{Hamidreza Payan is with the Department of Electrical Engineering, Amirkabir University of Technology, Tehran, Iran (e-mail:hamidrezapayan@aut.ac.ir).}
\thanks{Nima Alamatsaz is with the Department of Biomedical Engineering, New Jersey Institute of Technology, Newark, United States
 and also with the Graduate School of Biomedical Sciences, Rutgers University, Newark, United States (e-mail: nima.alamatsaz@njit.edu).}
\thanks{Fahimeh Nasimi is with the Department of Biomedical Engineering, Faculty of Engineering, University of Isfahan, Isfahan, Iran (e-mail: f.nasimi@eng.ui.ac.ir).}}

\maketitle

\begin{abstract}
Objective: Electrocardiogram (ECG) is the most frequent and routine diagnostic tool used for monitoring heart electrical signals and evaluating its functionality. The human heart can suffer from a variety of diseases, including cardiac arrhythmias. Arrhythmia is an irregular heart rhythm that in severe cases can lead to heart stroke and can be diagnosed via ECG recordings. Since early detection of cardiac arrhythmias is of great importance, computerized and automated classification and identification of these abnormal heart signals have received much attention for the past decades.
Methods: This paper introduces a light deep learning approach for high accuracy detection of 8 different cardiac arrhythmias and normal rhythm. To leverage deep learning method, resampling and baseline wander removal techniques are applied to ECG signals. In this study, 500 sample ECG segments were used as model inputs. The rhythm classification was done by an 11-layer network in an end-to-end manner without the need for hand-crafted manual feature extraction. Results: In order to evaluate the proposed technique, ECG signals are chosen from the two physionet databases, the MIT-BIH arrhythmia database and the long-term AF database. The proposed deep learning framework based on the combination of Convolutional Neural Network(CNN) and Long Short Term Memory (LSTM) showed promising results than most of the state-of-the-art methods. The proposed method reaches the mean diagnostic accuracy of 98.24\%. Conclusion: A trained model for arrhythmia classification using diverse ECG signals were successfully developed and tested. Significance: Since the present work uses a light classification technique with high diagnostic accuracy compared to other notable methods, it could successfully be implemented in holter monitor devices for arrhythmia detection.
\end{abstract}


\section{Introduction}\label{sec:intro}
Cardiovascular disease is amongst the three main global causes of death. As reported by the World Health Organization, nearly 18 million people die each year from heart disease \cite{world2002world}.\\
An electrocardiogram (ECG) is a method of measuring and recording the electrical activity of the heart. ECG signal is recorded non-invasively by placing multiple electrodes on particular points of the body. A normal heart rhythm consists of three main parts including P waves, T waves and QRS complex. The occurrence of any anomalies in the rhythm and/or heart rate indicates the existence of a disorder and is called arrhythmia \cite{mondejar2019heartbeat}. Heart arrhythmia is one of the chronic diseases that many people are suffering from worldwide. In some cases, arrhythmias lead to heart stroke or sudden cardiac death and therefore endanger human life. An early and accurate diagnosis of life-threatening arrhythmias can be effective in saving lives \cite{yildirim2018arrhythmia}.\\
People who are suffering from arrhythmias might have symptoms such as inadequate blood pumping, shortness of breath, fatigue, chest pain, and unconsciousness. However, its important to use a holter device for remote long-term heart monitoring for people with heart failure risk factors such as previous history in immediate family members or high blood pressure. Furthermore, for the analysis of long-term ECG recordings,
the manual examination is exhausting and time-consuming, especially in cases in which the real-time diagnosis is of matter. Therefore, in recent years, the use of automated computer-aided methods for diagnosis has been significantly increased and although several methods have been proposed, this matter still attracts the attention of scientists \cite{abdelmoneem2020arrhythmia}.\\
Recently, deep learning algorithms such as convolutional neural networks (CNN) as an automated computer-aided method have been frequently used for different biomedical tasks and yielded encouraging results \cite{tian2020deep}. The ability of deep learning networks to recognize patterns and learn important features from raw data makes them suitable for classifying ECG signals \cite{parvaneh2019cardiac}.\\
A quick search in the literature shows diverse methods proposed for arrhythmia classification. Most of the methods are based on heavy deep learning methods that use a high number of layers to detect arrhythmias. The high number of layers decelerates the training phase and later the evaluation phase resulting in sacrificing the real-time goal. Therefore, there is a need to develop a light deep learning model to classify ECG arrhythmias with high accuracy. \\
The purpose of this study is to develop an accurate light automatic diagnostic system that assists cardiologists by providing an intelligent deep learning method that is time-saving and cost-efficient and reduces the number of misdiagnosed arrhythmias. The proposed model deployed CNN and recurrent neural networks (RNN) such as long-term short memory (LSTM) for the detection of various cardiac arrhythmias.
 ECG recordings used in this study, were obtained from the MIT-BIH Arrhythmia Database and the Long-term AF database which are available from PhysioNet.As a summary, the main contributions and novelty of this paper are listed as follows:
\begin{enumerate}
\item LSTM along with CNN architectures has been deployed, because the prediction depends on the whole input sequence.
\item We focused on creating a lightweight model to automatically detect and classify 8 different arrhythmias as well as normal sinus rhythm. The arrhythmias that were discussed in this article are: atrial fibrillation, atrial flutter, ventricular bigeminy, Paced rhythm, Wolf Parkinson White syndrome (WPW), supraventricular tachyarrhythmia, ventricular trigeminy and ventricular tachycardia.
\end{enumerate}

The rest of the paper is organized as follows: Related works are presented in section \ref{RW}. The proposed system architecture is illustrated in section \ref{prp}. Experimental results are illustrated and then compared with results of other works in literature in section \ref{smrs}. Finally, in the conclusions section, the current condition of the suggested technique as well as the advantages and disadvantages of the work is represented.
\section{Literature review}\label{RW}
Several researchers have carried out different approaches throughout the years for classification and interpretation of various cardiac arrhythmias \cite{marinho2019novel,asl2008support,chen2020detection,sinha2020automatic}.
 Machine learning techniques as automated classification methods have boosted the classification accuracy in recent years. The study in \cite{aphale2021arrhynet} used ArrhyNet to detect and classify ECG signals and to overcome the issue of imbalanced dataset the utilized Synthetic Minority Over Sampling (SMOTE) technique. The accuracy of $92.73\%$ was achieved for their classification model.
Amongst numerous applied methods, deep learning techniques based on CNN models have been widely implemented due to their promising performance. One way of applying CNN models on ECG signals is by virtue of transfer learning. Authors in \cite{cimen22transfer}, proposed a perspective for accurate classification of rare arrhythmia types which are represented in 2-D image using transfer learning method, a pre-trained CNN model called VGG16 for extracting features and a v-SVM classifier for assorting. The accuracy of the their model for normal and arrhythmia signals is 87\% and 93\%, respectively, resulting in an average accuracy of 90.42\% .Isin et al. also used a transfer learning approach and a pre-trained CNN model called AlexNet as a feature extractor. Then a simple back-propagation was applied in order to classify the signals into three different cardiac rhythms (normal, Right Bundle Branch Block, paced). This transferred deep learning approach functioned efficiently by reaching an accuracy of 92\% on test dataset \cite{isin2017cardiac} . Mustaqeem et al.\cite{mustaqeem2018multiclass} proposed a method for differentiating between normal and diseased ECG signals obtained from the UCI database. The proposed approach for the selection of the most significant features of the ECG signals was a wrapper method, built on a random forest algorithm. Afterwards, cardiac arrhythmias were classified by means of three different SVM based techniques including one-against-all (OAA), one-against-one (OAO), and error-correcting codes (ECC). After evaluating the mentioned techniques, the OAO method proved to be best suited for classifying the ECG records into 16 categories, by achieving an accuracy of 92.07\% when using a 90:10 data split ratio . Hannun et al.\cite{hannun2019cardiologist} developed a 34-layer end-to-end deep neural network for categorizing 12 rhythm classes. The DNN model was conducted on a novel vast single-lead ECG database and resulted in an encouraging AUC of 0.97. Authors in\cite{ilbeigipour2021real} executed a segmentation operation on the ECGs existing in the MIT-BIH database.They broke down the large records into segments containing 200 samples. Subsequently, three decision trees, random forest and logistic regression multi-class classifiers were used in order to classify 3 cardiac rhythms. The average rhythm detection time is about 1 second resulting in online and real-time performance. Eventually, random forest showed more accurate and stronger performance by achieving 88.7\% accuracy and 92.5 precision.
Authors in\cite{lu2021kecnet} used KecNet model, which contains a CNN structure with a modified convolutional layer and a symbolic parameter extraction architecture in the feature extraction part of the model to classify arrhythmias. They achieved the accuracy of $99.31\%$. Authors in \cite{kanani2020ecg} augmented a set of time-series transformations of the original signals to the original dataset and used a 1-D CNN model to classify arrhythmias using the original and transformed database.They managed to attain an accuracy of $99\%$ without overfitting the model.
\section{Arrhythmia classification methodology}\label{prp}
In the arrhythmia detection and classification method proposed in this study, first a pre-processing is done to make the data ready to fed into deep neural network algorithm and at the end the training and testing procedure using deep neural network algorithm is accomplished on data . The pre-processing section is made up of following steps: noise elimination and data resampling. The research is carried out in Python programming language in JupyterLab from the Anaconda distribution of Python 3.8.11 on a system with Intel Core i7 7th Gen processor with NVIDIA GeForce GTX 1070 Ti using TensorFlow and TensorFlow-GPU  2.3.0 packages.
\subsection{ECG Dataset}
In this study, the MIT-BIH arrhythmia database and the long-term atrial fibrillation (LTAF) database were used for train and test purposes of the proposed framework for arrhythmia classification \cite{goldberger2000physiobank}.
The MIT-BIH arrhythmia database consists of 48 half-hour ECG recordings that were obtained from 47 subjects, studied by the Beth Israel Hospital Arrhythmia Laboratory. Unlike the conventional ECG recordings that are recorded by $12$ leads, this database contains heart signals which were recorded by two leads (usually  ML\RNum{2} and V1 leads) with a $\SI{360}{\hertz}$ sampling frequency\cite{moody2001impact}.
The LTAF database includes 24-hours ECG signals that were obtained from $84$ subjects with paroxysmal atrial fibrillation. ECG signals were recorded synchronously from two leads with $\SI{128}{\hertz}$ sampling frequencies \cite{petrutiu2007abrupt}. These ECG recordings were annotated with details such as rhythm type, beat type, peak locations and onset and offset of a waveform by two or more professional cardiologists. These annotations were first extracted from the signals and then used in the train and test process. Each subject's ECG recordings may contain different types of arrhythmias, so by using the rhythm type annotations, all arrhythmias were excerpted from all subjects.
\subsection{Preprocessing}
 \subsubsection{Noise filtering}
ECG signals are usually corrupted by different types of either low- or high- frequency noises such as baseline wander (BW), power line interference, electromyography (EMG) noise and electrode motion artifact noise. Different types of filters can be applied to remove these noises. Respiration and subjects movement are the main cause of BW which is a low-frequency artefact in ECG signal recordings of a subject. In this research, following previous works, a median filter with 200 ms and 600 ms width is used \cite{mondejar2019heartbeat}. The median filter is a non-linear digital filtering technique used for noise removal of images and signals whilst preserving the useful details of the signal or the image. Each record was then normalized to an amplitude range of $[-1,+1]$.
\subsubsection{Resampling}
The MIT-BIH ECG recordings were digitized at $360$ samples per second and the ECG recordings in the LTAF database were digitized at $128$ samples per second. Therefore in order to make use of both databases a resampling technique is being use to downsample the signals in  MIT-BIH dataset. So after the resampling process the frequency of all the records are $\SI{128}{\hertz}$.
\subsection{Segmentation}
Segmentation of the ECG recordings is a step toward homogenizing the length of the data that is going to be fed to the model. With the sampling rate of $\SI{128}{\hertz}$ and an average cardiac cycle of $0.8$ second, segments with $500$ samples ($3.9$ seconds) seem appropriate, since most arrhythmias appear within this length. Segments were extracted in an overlapping manner. The segmenting window slides through the records and produces the sections.
After this step, all ECG segments from both databases are combined together. As shown in \autoref{tab1} segments related to normal and atrial fibrillation classes were excessively high. Therefore in order to eliminate the side effects of this imbalance, evaluation metrics during both the training and testing stages were weighted by the inverse of the size of each class.

\begin{table}[!h]
\caption{Number of each rhythm type before and after segmentation}
\setlength{\tabcolsep}{3pt}
\begin{tabular}{|p{55pt}|p{60pt}|p{60pt}|p{50pt}|}
\hline
Arrhythmia type&	Num of rhythms in MIT-BIH&	Num of rhythms in LTAF&	Total num of segments\\
\hline
Atrial fibrillation (AFIB)&	107	&7358&	10000\\
\hline
Atrial flutter (AFL)	&45&	-&	1099\\
\hline
Ventricular bigeminy (B)	&221&	2696	&10687\\
\hline
Normal rhythm (N)&	530	&22834&	10000\\
\hline
Paced rhythm(P)	&60	&-&	7680\\
\hline
Wolf-Parkinson-white syndrome (PREX)&	103&	-&	525\\
\hline
Superventricular tachycardia (SVTA)&	26&	3268&	6987\\
\hline
Ventricular trigeminy (T)	&83	&785	&4752\\
\hline
Ventricular tachycardia (VT)&	61&	824	&198\\
\hline

\end{tabular}
\label{tab1}
\end{table}
\subsection{Proposed model architecture}
This paper aims to introduce a high-accuracy deep learning technique based on the combination of CNN and LSTM architectures, to diagnose different types of cardiac arrhythmias in an end to end manner from raw ECG signals. In this study the proposed model consists of $11$ layers, which are trained using the $85\%$ of the data and is evaluated using the other $15\%$ of the segmented data.

\begin{figure}[!t]
\centerline{\includegraphics[width=\columnwidth]{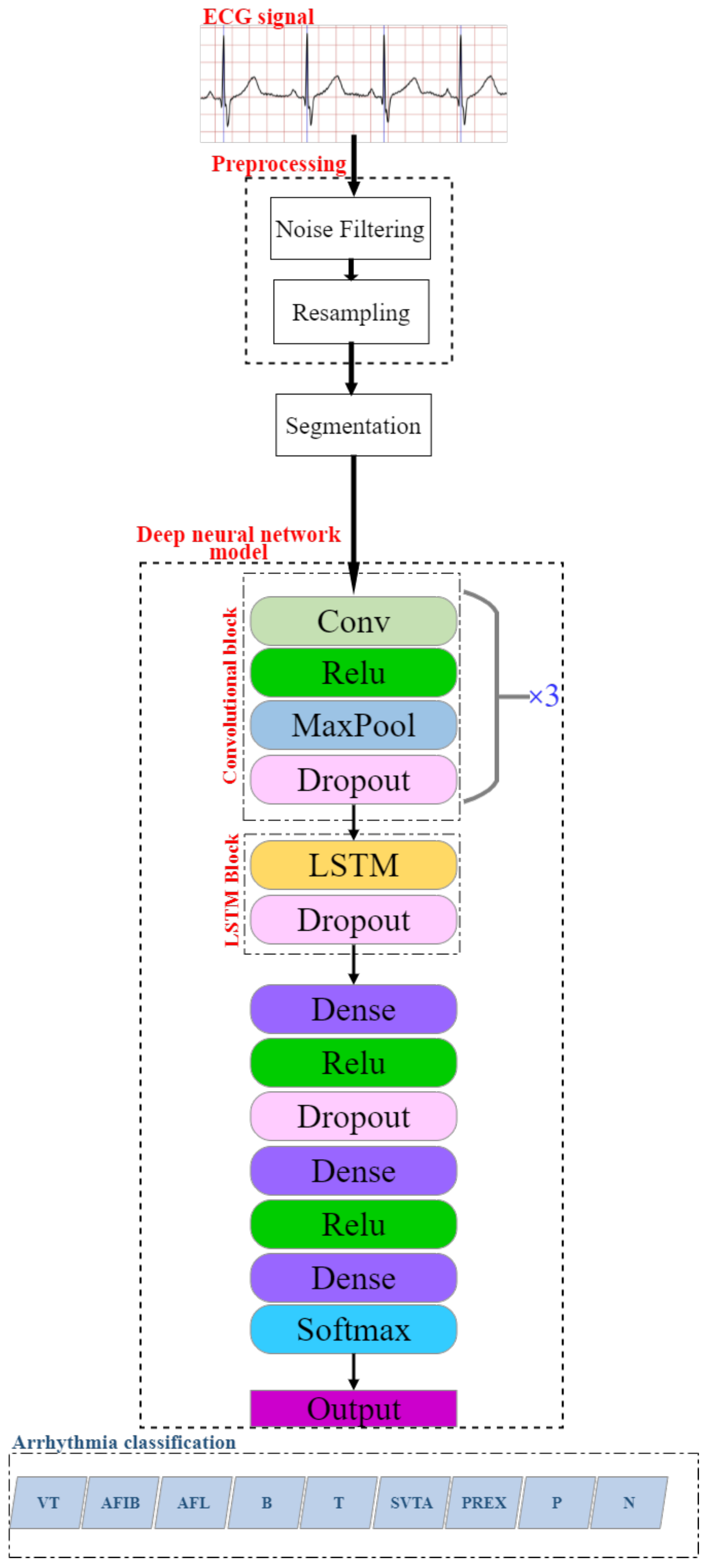}}
\caption{Architecture of the proposed deep learning model}
\label{fig1}
\end{figure}

The architecture of the proposed model as illustrated in \autoref{fig1} is as follows: three tandem convolutional blocks that each one contains a 1-dimensional convolutional layer for extracting high-level features and an activation function, called Rectified Linear Unit (ReLu) for achieving non-linear capabilities. The output of Relu layers is then given to the max-pooling layer for reducing feature dimensions and selecting the most significant features. The output of the last convolutional block is then fed to a LSTM layer. LSTM consists of memory blocks and has a recursive feedback connection that can handle long-term dependencies and exploding gradient problems. In order to prevent the proposed deep neural network model from overfitting, dropout layers are used. This layer by randomly dropping out a fraction of nodes prevents the overfitting problem.
Fully connected layers usually form the last few layers of the network configuration. In this model, several dense layers were used for changing the vector’s dimensions and altering the few-dimensional features to linear vectors. As the last layer multi-class activation function, softmax, was used for converting the output vector of classes to probabilities for each class. Since the target variable is categorical, one-hot encoding process were used during the multi-classification procedure. \autoref{tab2} gives the detailed description of the CNN model joined by a LSTM layer implemented for this study.

\section{Results and discussion}\label{smrs}
The purpose of this research is to classify 8 arrhythmias and normal rhythm. Therefore after data preparation, the dataset was divided into two groups of training and testing. Experiments confirm that the more the size of training data, the better its functionality on test data, so 85\% of the prepared data has been randomly chosen and used for training the proposed model and 15\% of the data has been used for testing and validating the model. The model was compiled for 100 epochs in the training and testing set.
Categorical cross-entropy was used as the loss function and Adam optimizer was chosen because it makes the algorithm convergence faster, its implementation is straightforward, it is computationally efficient and its default hyper-parameters can be used with little tunning. Weights and biases are updated continuously using a conjugate gradient back-propagation algorithm which utilizes the generated value by the loss function until the desired optimized values are acquired.\\
\subsection{Performance evaluation metrics}
There are several different metrics to assess the performance of a multi-class classification model. Three of which are used in this study are accuracy (Acc), sensitivity (Se), and specificity (Sp).\\
As shown in \eqref{eq1} accuracy, indicates how accurate the model has performed
\begin{equation}
Acc(\%)\triangleq\frac{TP+TN}{TP+TN+FP+FN}\times100\%.
\label{eq1}
\end{equation}
Sensitivity is the metric that evaluates a models ability to predict true positives of each class.
\begin{equation}
Se(\%)\triangleq\frac{TP}{TP+FN}\times100\%.
\label{eq2}
\end{equation}
Specificity is the metric that evaluates a models ability to predict true negatives of each class.
\begin{equation}
Sp(\%) \triangleq \frac{TN}{TN+FP}\times100\%.
\label{eq3}
\end{equation}

\begin{table}[!h]
\caption{Detailed description of the proposed deep learning model }
\setlength{\tabcolsep}{3pt}
\begin{tabular}{|p{50pt}|p{42pt}|p{42pt}|p{42pt}|p{42pt}|}
\hline
Layers	&Filter size&	Kernel size	&pool size&	strides\\
\hline
Conv1D&	64&	50&	-	&-\\
\hline
MaxPool1D&-&	-	&20&	2\\
\hline
Dropout&	 \multicolumn{4}{|c|}{Dropout rate = 0.1}\\
\hline
Conv1D	&32&	10&	-	&-\\
\hline
MaxPool1D&	-&-	&10&	2\\
\hline
Dropout&\multicolumn{4}{|c|}{	Dropout rate = 0.1}\\
\hline
Conv1D	&16&	5&	-&	-\\
\hline
MaxPool1D	&-&	-&	5&	2\\
\hline
Dropout&	\multicolumn{4}{|c|}{Dropout rate = 0.1}\\
\hline
LSTM	&\multicolumn{4}{|c|}{units=32}\\
\hline
Dense	&\multicolumn{4}{|c|}{units=32}\\
\hline
Dropout	&\multicolumn{4}{|c|}{Dropout rate = 0.1}\\
\hline
Dense	&\multicolumn{4}{|c|}{units=16}\\
\hline
softmax &\multicolumn{4}{|c|}{output=9}\\
\hline
\end{tabular}
\label{tab2}
\end{table}
Above, true positive (TP) means segments in which the true label is positive and whose class is correctly predicted to be positive whereas false positive (FP) means segments which the true label is negative and whose class is incorrectly predicted to be positive. True negative (TN) shows segments which the true label is negative and whose class is correctly predicted to be negative whilst false negative (FN) shows segments which the true label is positive and whose class is incorrectly predicted to be negative.
\subsection{Experimental results and discussion }
\begin{figure}[!h]
\centerline{\includegraphics[width=\columnwidth]{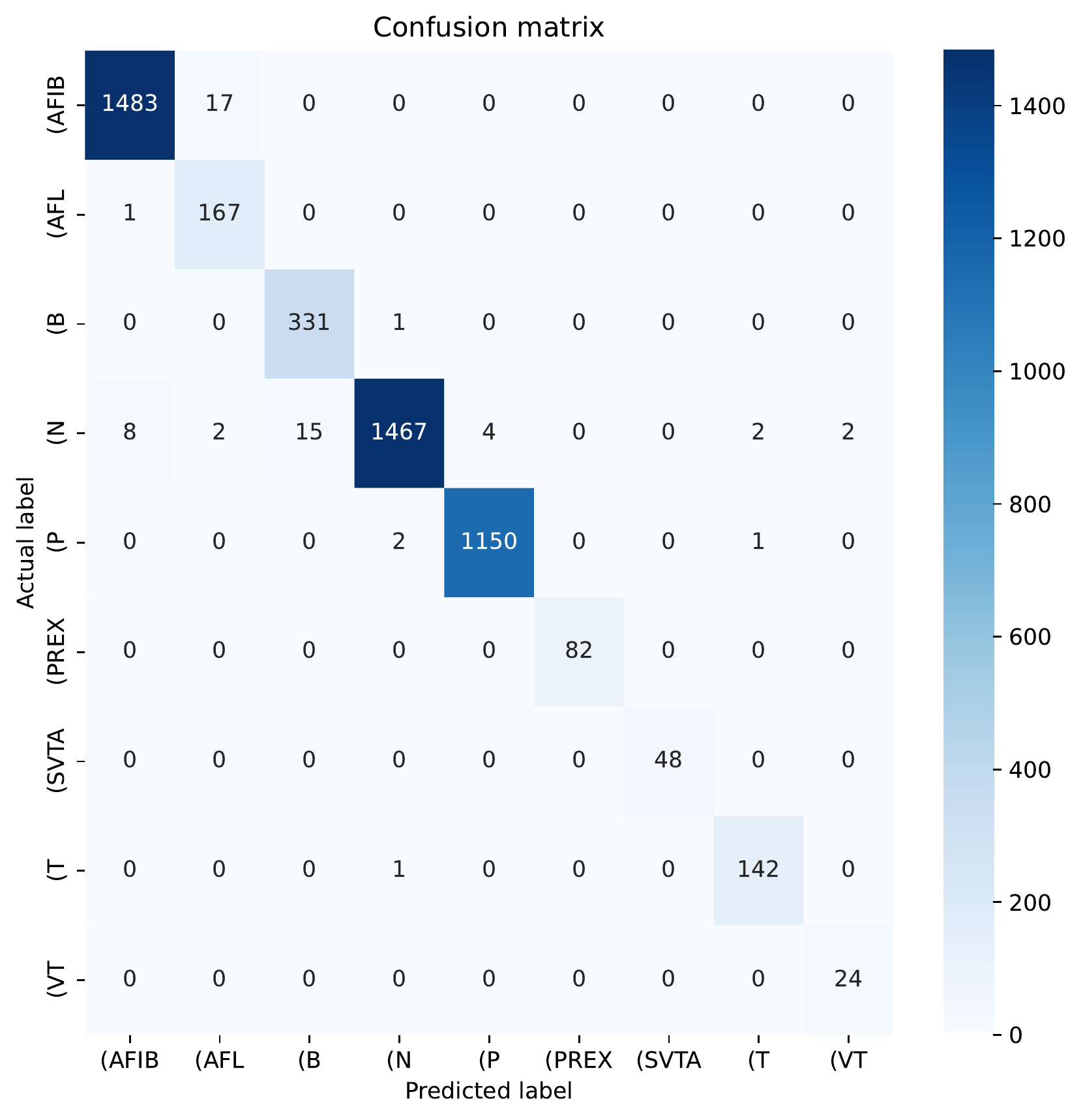}}
\caption{Confusion matrix for the predictions of the proposed method}
\label{fig2}
\end{figure}

\autoref{fig2} gives the confusion matrix that is used to visualize the performance of the classification algorithm. Numbers on the main diameter of this matrix indicate true positives.

\begin{figure}[!h]
\centerline{\includegraphics[width=\columnwidth]{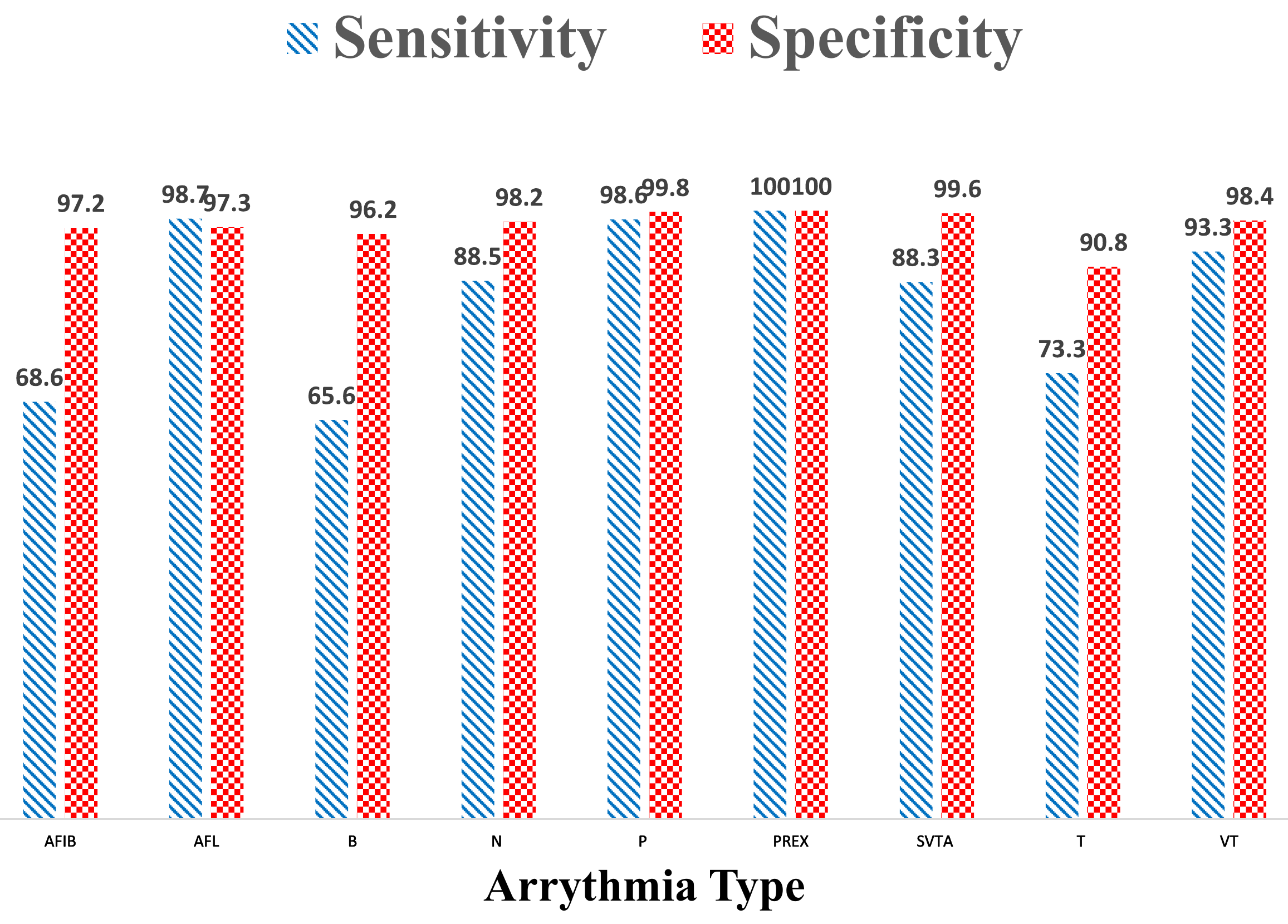}}
\caption{Evaluation results of the proposed technique}
\label{spse}
\end{figure}

Sensitivity and specificity parameters which are independent of the number of segments, are used to evaluate the proposed technique. As shown in \autoref{spse} the ability of the proposed model to correctly identify other rhythms when a particular rhythm is considered (specificity) is above 90\% for all 8 arrhythmias and normal rhythm.\\
A quick review on false negative rates of AFIB, B and T arrhythmias shown in \autoref{fig2} unveiled that, this poor identification of the desired arrhythmias entirely appear very sensible. In many cases, the lack of context, limited signal duration, or having a single lead limited the derivation that could be concluded from the data, making it challenging to certainly reveal whether the annotating cardiologists and/or the algorithm was correct.\\
Accuracy indicates how accurate the model has performed. The average classification accuracy achieved by the proposed model was 98.24\%. The novel 1D-CNN+LSTM model implemented in this study exhibited a high-performance accuracy for the classification of different arrhythmia types. Its implementation is straightforward and has lower computational complexity in comparison with most of the state-of-the-art approaches such as SVM classifiers-based strategies, random forest algorithm \cite{mustaqeem2018multiclass} or deployment of ensemble classifiers alongside SVM-based methods. Besides, the proposed model consists of a relatively low number of layers contrary to the models that were executed in \cite{hannun2019cardiologist,lu2021kecnet,aphale2021arrhynet,kanani2020ecg}. Most previous researches used only one database of ECG signals \cite{ilbeigipour2021real,yildirim2018arrhythmia} but a combination of two databases with different sampling frequencies was used for training and testing procedures. Furthermore, only a limited number of arrhythmia disorders were classified in  most of earlier studies such as \cite{isin2017cardiac,ilbeigipour2021real} but 9 different cardiac rhythm types were distinguished in this system.The performance comparison with other recent research studies about the same problem is given in \autoref{tab3}.

\begin{figure*}[!h]
\centerline{\includegraphics[width=\textwidth,height=9cm]{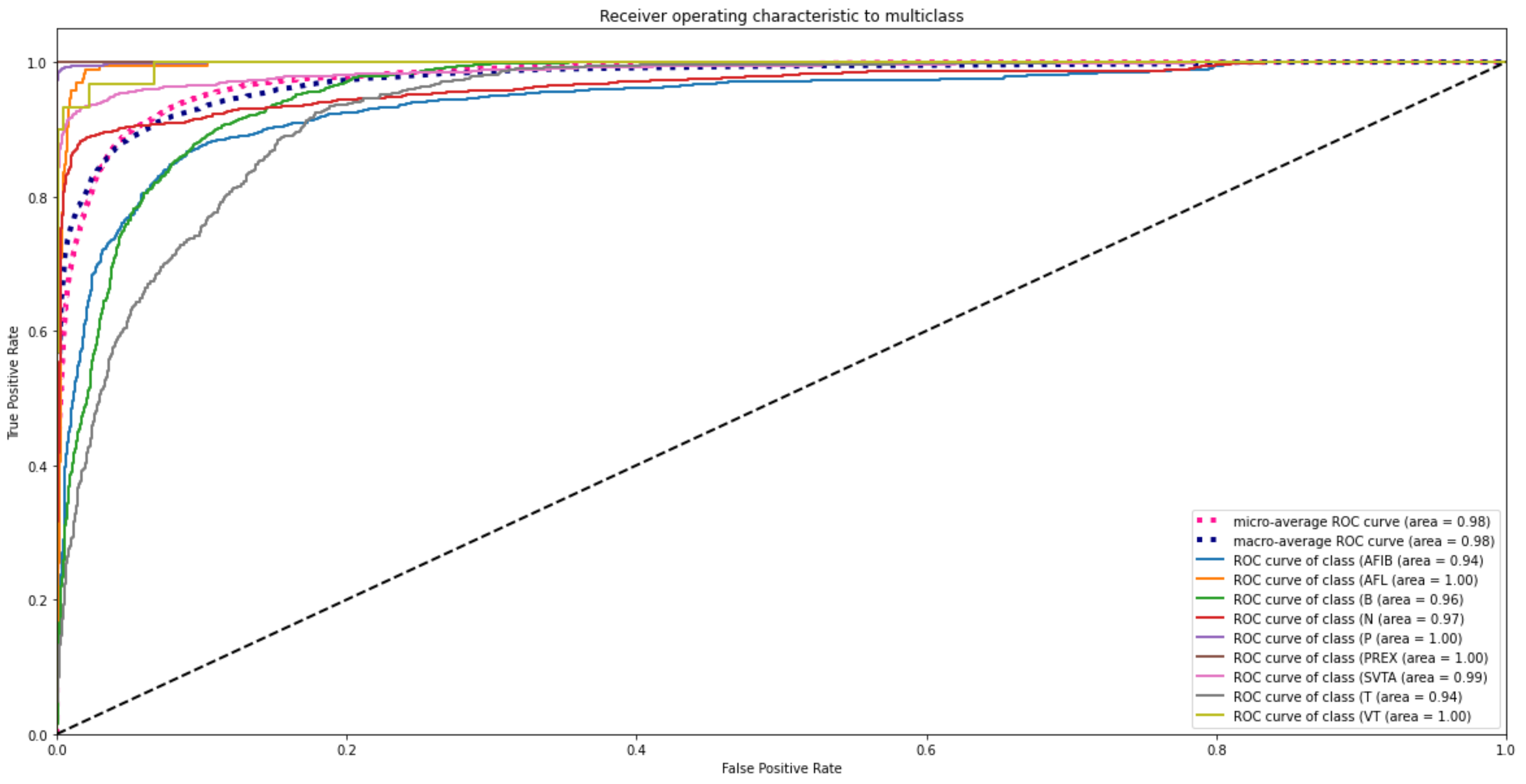}}
\caption{ Receiver operating characteristic curves for proposed model predictions on 9 rhythm classes.}
\label{roc}
\end{figure*}

ROC curve is an evaluation metric that gives a graphical illustration of a classifier diagnostic capability. It is actually generated by plotting true positive rate (TPR) also known as sensitivity against false positive rate (FPR) also known as (1-specificity) at different threshold settings. The Area Under the Curve (AUC) is the measure that shows how well the classifier has discriminated between classes. Hence, the closest the AUC is to 1, the better the performance of the model. As it is shown in \autoref{roc}, the model reached nearly perfect AUC in distinguishing between each arrhythmia class. \\

\begin{table}[!h]
\caption{ Summary of automated arrhythmia detection techniques}
\setlength{\tabcolsep}{3pt}
\begin{tabular}{|p{70pt}|p{62pt}|p{30pt}|p{30pt}|p{20pt}|}
\hline
Ref(work)&	Classification technique&	Number of Layers&	accuracy	&Num of classes\\
\hline
P.Lu et al.\cite{lu2021kecnet}&	Light Neural Network&	15	&99.31\%&	5\\\hline
S.Aphale et al.\cite{aphale2021arrhynet}&	CNN(ArrhyNet)&15 &	92.73\%&	5\\\hline
A.Isin et al.\cite{isin2017cardiac}&	Transferred deep CNN&	8	&92\%	&3\\\hline
P.kanani et al.\cite{kanani2020ecg}&	Time series and Deep learning &	22&	99.12\%&	5\\\hline
S.Ilbeigipour et al.\cite{ilbeigipour2021real}&Structured Streaming Module&	-&	88.7\%&3\\\hline
A.Mustaqeem et al.\cite{mustaqeem2018multiclass}&Support Vector Machine(SVM)	&-	&92.07\%&	16\\\hline
A.Hannun et al.\cite{hannun2019cardiologist}	&DNN&	34&85\%&12\\\hline
E.cimen et al.\cite{cimen22transfer}&Transfer Learning&	18&	90.42\%&2\\\hline
\textbf{Proposed}&	1D-CNN + LSTM&	11&	98.24\%&	9\\

\hline
\end{tabular}
\label{tab3}
\end{table}
The high computational complexity of CNNs presents a critical challenge towards their broader adoption in real-time and power-efficient scenarios. In the following, in order to prove that our model is lightweight and could be used in holter monitor devices, we will show the model size and inference time to classify a single rhythm.
The inference time of the proposed model on raspberry pi is only $ \SI{5.127}{ \ms}$, meaning that it takes $ \SI{5.127}{ \ms}$ to classify one rhythm. The size of model being loaded on the Raspberry pi(processor being used in holter monitors) is $ \SI{0.16}{\mega\byte}$ .This implies that our model is competitive on a Raspberry pi-based holter monitor devices.
\section{Conclusion and Future work}
Correct detection of cardiac arrhythmias is crucial for early treatment of the patients and computer-aided diagnosis can play an important role. In this paper, the experiment has been conducted on ECG recordings obtained from the two databases MIT-BIH and Long-term AF. The proposed CNN+LSTM model is able to perform the classification of 8 different types of arrhythmias and normal ECG signals. The model can extract discriminant features and information of the heart signals through CNNs and temporal features through a LSTM layer. The experiment reached an average testing accuracy of 98.24\% with the inference time of $ \SI{5.127}{ \ms}$ for a single unseen rhythm. \\
In addition to using the shape of the heartbeat to detect arrhythmia, utilising the other changes in the property of the signal for training the model could be beneficial to achieving better performance. Some of these attributes are RR intervals and QRS intervals (period).
For this model, convolution neural networks were followed by an LSTM layer to gain better accuracy in arrhythmia detection. However, the classification accuracy of some of the arrhythmias still needs improving. Deploying some rule-based algorithms alongside this model can lead to better performance.
Most of the previous research has been conducted on only one database (mostly MIT-BIH) but this research was based on the combination of two databases. However, the combination of more databases is required for more reliable results.
 The ECG signals used in this study for training and testing the model were obtained by two leads (usually MLII and V1 leads), while in clinical applications 12-lead ECGs are known as standard. An ideal model should be able to discriminate against the standard ECG signals.
Although the model was able to classify 8 types of cardiac arrhythmias as well as the normal sinus rhythm with high accuracy, which is more than what some of the related works have been capable of, there still are various important undetected arrhythmias and heart disorders. The implemented model should be improved to be able to distinguish between more arrhythmias.


\bibliographystyle{IEEEtran}
\bibliography{references}

\end{document}